\documentclass{ws-procs9x6}

\begin{document}

\title{
Deformation and weak decay of $\Lambda$ hypernuclei}

\author{K. Hagino, Myaing Thi Win, and Y. Nakagawa}

\address{
Department of Physics, Tohoku University, Sendai, 980-8578, Japan}

\begin{abstract}
We use the self-consistent mean-field theory to discuss the 
ground state and decay properties of $\Lambda$ hypernuclei. 
We first discuss the deformation of $\Lambda$ hypernuclei using the 
relativistic mean-field (RMF) approach. We show that, 
although 
most of hypernuclei have a similar deformation
parameter to the core nucleus,
the shape of $^{28}$Si is drastically 
altered, from oblately deformed to spherical, 
if a $\Lambda$ particle is added to this nucleus. 
We then discuss the pionic weak decay of neutron-rich 
$\Lambda$ hypernuclei using the Skyrme Hartree-Fock + BCS method. 
We show that,
for a given isotope chain,
the decay rate increases as a function of mass number, due to the
strong neutron-proton interaction.
\end{abstract}


\bodymatter

\section{Introduction}

The self-consistent mean-field method has been 
a standard approach to the description of ground state properties of 
atomic nuclei\cite{BHR03}. 
It has been extensively applied also to 
hypernuclei\cite{R81,RSM90,LY97,MJ94}, in which 
the mass number dependence of $\Lambda$ binding energy has 
been successfully reproduced, from a 
light nucleus $^{12}_{~\Lambda}$C 
to a heavy nucleus $^{208}_{~~\Lambda}$Pb. 

In this contribution, we employ the self-consistent mean-field approach 
to discuss the structure and decay of $\Lambda$ hypernuclei. 
In the first part, we discuss the deformation of 
$\Lambda$ hypernuclei \cite{MH08}. 
It has been well known that many open-shell nuclei are deformed in the 
ground state. 
By allowing the rotational symmetry to be broken in the mean-field potential, 
the mean-field theory provides an intuitive and transparent view of the 
nuclear deformation \cite{BHR03}. 
Recently, the deformation property of $\Lambda$ hypernuclei has 
been explored in a broad 
mass region using the 
non-relativistic Skyrme Hartree-Fock method \cite{ZSSWZ07}. 
Here we carry out a similar study using 
the relativistic mean field (RMF) method   
as an alternative choice of effective $NN$ and $N\Lambda$ interactions. 

In the second part, we discuss the weak decay of neutron-rich 
$\Lambda$ hypernuclei. 
Neutron-rich nuclei are one of the current topics in nuclear physics. 
In connection to hypernuclear physics, there have been discussions 
on an extension of neutron-drip line as a consequence of an additional 
$\Lambda$ particle \cite{VPLR98,ZPSV08}. 
It has been well 
known that, in finite nuclei, the mesonic (pionic) decay, 
$\Lambda\to N\pi$, 
is largely Pauli-suppressed as the mass 
number increases, and the non-mesonic decay, 
$\Lambda N\to NN$, is the dominant decay mode \cite{C90,S05}. 
However, it has not been known well how the pionic decay mode is suppressed 
as the neutron number increases for a fixed value of proton number. 
Here we address this question for relatively light neutron-rich 
hypernuclei
using the non-relativistic 
Skyrme Hartree-Fock method. 

\section{Deformation of $\Lambda$ hypernuclei}

Let us first discuss the deformation of $\Lambda$ hypernuclei \cite{MH08}. 
In the RMF approach, nucleons and a $\Lambda$ particle are treated as 
structureless Dirac particles, interacting through the exchange of virtual 
mesons, that is, the isoscalar scalar $\sigma$ meson, 
the isoscalar vector $\omega$ meson, and the isovector vector $\rho$ meson. 
The photon field is also taken into account to describe the Coulomb 
interaction between protons. 
The effective Lagrangian for $\Lambda$ hypernuclei may be given 
as \cite{RSM90,VPLR98} 
\begin{equation}
{\cal L}={\cal L}_N+\bar{\psi}_\Lambda\left[
\gamma_\mu\left(i\partial^\mu-g_{\omega\Lambda}\omega^\mu\right)
-m_\Lambda-g_{\sigma\Lambda}\sigma\right]\psi_\Lambda,
\label{RMF}
\end{equation}
where $\psi_\Lambda$ and 
$m_\Lambda$ are 
the Dirac spinor and the mass for the $\Lambda$ particle, respectively. 
Notice that the $\Lambda$ particle couples only to the 
$\sigma$ and $\omega$ mesons, as it is neutral and isoscalar. 
${\cal L}_N$ in Eq. (\ref{RMF}) is the standard RMF Lagrangian for the 
nucleons. 

We solve the RMF Lagrangian (\ref{RMF}) in the mean field approximation. 
The variational principle leads to the Dirac equations for the nucleons 
and the $\Lambda$ 
particle, and the Klein-Gordon equation for the mesons. 
We solve these equations iteratively until the self-consistency condition is 
achieved. 
For this purpose, we 
modify the computer code {\tt RMFAXIAL} \cite{RGL97} 
to include the $\Lambda$ particle. 
The pairing correlation among the nucleons is also  
taken into account in the constant gap approximation. 

With the self-consistent solution of the RMF equations, we compute the 
intrinsic quadrupole moment of the hypernucleus,
\begin{equation}
Q=\sqrt{\frac{16\pi}{5}}\,\int d\vec{r}\,[\rho_v(\vec{r})+
\psi_\Lambda^\dagger(\vec{r})\psi_\Lambda(\vec{r})]\,r^2Y_{20}(\hat{\vec{r}}), 
\end{equation}
from which we estimate 
the quadrupole deformation parameter $\beta_2$. 

\begin{figure}
\begin{center}
\psfig{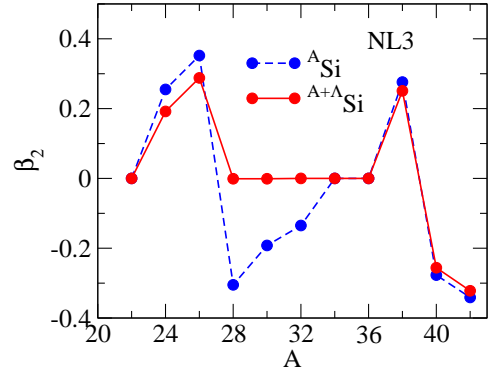}
\end{center}
\caption{
Quadrupole deformation parameter for Si isotopes 
obtained with the RMF method with the NL3 parameter set. 
The dashed line is the deformation parameter for the core nucleus, while 
the solid line is for the corresponding hypernucleus. 
}
\end{figure}

\begin{figure}
\begin{center}
\begin{tabular}{cc}
{
\hspace*{-2cm}
\psfig{file=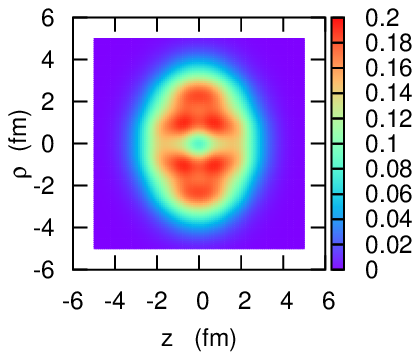,width=3in,clip}}&
{
\hspace*{-2cm}
\psfig{file=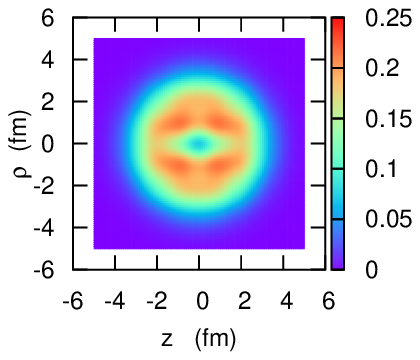,width=3in,clip}}
\end{tabular}
\end{center}
\caption{
The density profile for the $^{28}$Si (the left panel) and 
for the $^{29}_{~\Lambda}$Si (the right panel) obtained with the RMF method. 
It is plotted 
with the cylindrical coordinate $(z,\rho)$, where 
the $z$ axis is the symmetry axis. 
The unit for the density is fm$^{-3}$. 
}
\end{figure}

Figure 1 shows the deformation parameter for the ground state 
of Si isotopes obtained with the NL3 parameter set\cite{nl3}. 
The dashed line is the deformation 
parameter for the even-even core nuclei, while the solid line is for 
the corresponding hypernuclei. 
We see that the deformation parameter for the 
$^{28,30,32}$Si nuclei is drastically changed when a $\Lambda$ particle is 
added, although the change for the other Si isotopes is small. 
That is, the $^{28,30,32}$Si nuclei have oblate shape in the ground state. 
When a $\Lambda$ particle is added to these nuclei, remarkably they 
turn to be spherical. The corresponding density profile for 
$^{28,28+\Lambda}$Si is shown in Fig. 2. 
The potential energy surfaces for the 
$^{28,28+\Lambda}$Si nuclei are shown in Fig. 3. 
The energy surface for the $^{28}$Si nucleus shows a 
relatively shallow oblate minimum, 
with a shoulder at the spherical configuration. 
The energy difference between the oblate and the spherical configurations 
is 0.754 MeV, and could be easily inverted when a $\Lambda$ particle is 
added. 

\begin{figure}
\begin{center}
\psfig{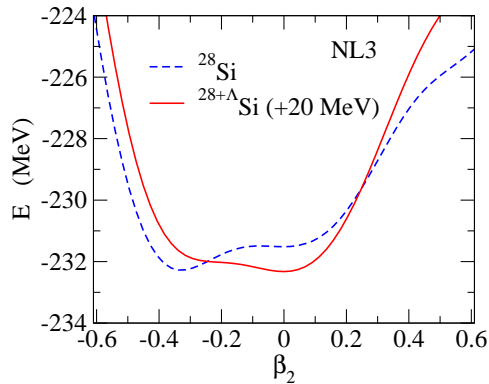}
\end{center}
\caption{
The potential energy surface for the $^{28}$Si (the dashed line) 
and $^{28+\Lambda}$Si (the solid line) nuclei obtained with the 
constrained RMF method with the NL3 parameter set. 
The energy surface for $^{28+\Lambda}$Si is shifted by a constant 
amount as indicated in the figure. 
}
\end{figure}


\section{Mesonic decay of neutron-rich $\Lambda$ hypernuclei}

Let us next discuss the pionic decay of neutron-rich $\Lambda$ hypernuclei. 
To this end, we use the formalism given in Ref. \cite{BT84}, where 
the standard Hamiltonian for the pionic decay of $\Lambda$ particle 
is evaluated with single-particle wave functions obtained with a 
mean-field approximation. 
Here, we use the Skyrme-Hartree-Fock method with SIII parameter 
set\cite{SIII}. As for the $\Lambda N$ interaction, we use the parameter 
set No. 1 given in Ref. \cite{R81}. The pairing among 
nucleons is taken into account in the BCS approximation with a 
density-dependent contact interaction. Continuum states are discretized 
within a large box. 

\begin{figure}
\begin{center}
\psfig{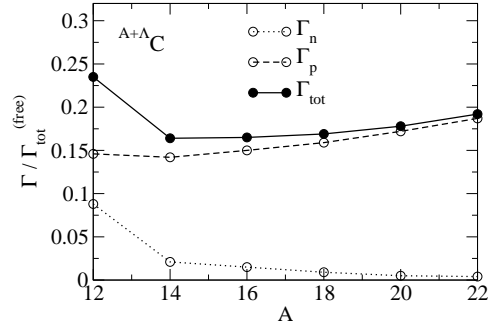}
\end{center}
\caption{
The pionic decay width for C hypernuclei obtained with the Skyrme 
Hartree-Fock method. The dotted and dashed lines are 
for the neutron and the proton modes, respectively, while the solid 
line shows the total width. These are plotted in unit of 
the total decay width of a free $\Lambda$ 
particle, $\Gamma_{\rm tot}^{\rm (free)}$=2.50$\times 10^{-12}$ MeV. 
}
\end{figure}

Figure 4 shows the pionic decay width for C isotopes obtained in this way. 
The final state interaction of pion is not taken into account. 
The dashed and dotted lines show the decay width for the proton 
($\Lambda\to p +\pi^-$) and the neutron 
($\Lambda\to n +\pi^0$) 
modes, respectively. The total width is denoted by the solid line. 
These are plotted in unit of 
the total decay width of a free $\Lambda$ 
particle, $\Gamma_{\rm tot}^{\rm (free)}$=2.50$\times 10^{-12}$ MeV. 
One sees that the decay width for the neutron mode is suppressed 
due to the Pauli principle, as expected. On the other hand, that for 
the proton mode is increased in neutron-rich nuclei. 
A similar conclusion has been obtained also with shell model 
calculations\cite{MI94}. 

\begin{figure}
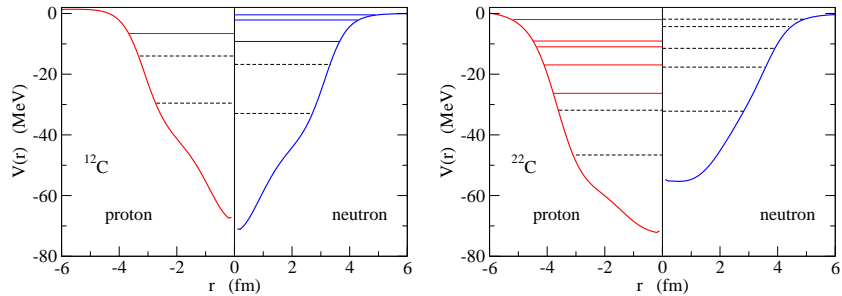

\begin{center}
\begin{tabular}{cc}
{
\hspace*{-1cm}
\psfig{file=fig5-1.eps,width=2.1in,clip}}&
{
\hspace*{-0.3cm}
\psfig{file=fig5-2.eps,width=2.1in,clip}}
\end{tabular}
\end{center}
\caption{
The single-particle potentials for $^{12}$C (the left panel) and 
for $^{22}$C (the right panel). For each panel, 
the proton well is shown on the left hand side, while the neutron well 
is shown on the right hand side. The occupied levels are denoted by the 
dotted lines, while the unoccupied levels are denoted by the solid line. 
}
\end{figure}

This behaviour can be understood if we consider a change in the mean-field 
potential. Fig. 5 shows the single-particle potentials for $^{12,22}$C. 
The occupied and the unoccupied levels are denoted by the dotted and the solid 
lines, respectively. As the number of neutron increases, the proton 
mean-field potential is deepened whereas the neutron mean-field potential 
becomes shallower. This is because of a strong neutron-proton interaction. 
As the proton single-particle potential well is deep, the number of 
bound (unoccupied) levels increases. At the same time, the momentum of the 
emitted pion increases. Both of these facts result in the enhancement of 
pionic decay width in neutron-rich nuclei, as shown in Fig. 4. 

\section{Summary}

We have first used the relativistic mean field (RMF) theory to investigate
quadrupole deformation of $\Lambda$ hypernuclei.
We have shown that, while an addition of $\Lambda$ particle does not
influence much the shape of many nuclei,
$^{28}$Si makes an important exception.
That is, we have demonstrated that
the $\Lambda$ particle makes the shape of this nucleus 
change from oblate to spherical.
In the second part, we investigated the pionic 
decay of neutron-rich $\Lambda$ hypernuclei. 
We have shown that the neutron mode is largely Pauli-suppressed 
as the number of neutron increases, while the decay width for the 
proton mode increases. We have argued that this is because the proton 
mean-field potential is deepened in neutron-rich nuclei. 

\section*{Acknowledgments}

We thank H. Tamura, T. Koike, H. Sagawa, H.-J. Schulze, and T. Motoba for
useful discussions.
This work was supported by the Japanese
Ministry of Education, Culture, Sports, Science and Technology
by Grant-in-Aid for Scientific Research under
the program number 19740115.

\end{document}